\documentclass[]{tPHM2e}

\usepackage{graphicx}% Include figure files 
\usepackage{color} 
\usepackage{psfrag}
 \newcommand{\ket}[1]{\left| #1 \right \rangle } 
\newcommand{\hc}{\hat{c}^{\phantom{\dagger}}}
\newcommand{\hcd}{\hat{c}^{\dagger}}
\newcommand{\hh}{\hat{h}^{\phantom{\dagger}}}
\newcommand{\hhd}{\hat{h}^{\dagger}}
\newcommand{\langind}[1]{\hspace{-0.1cm}{\phantom{\rangle}}_{#1}
\hspace{-0.05cm} \langle\hspace{0.03cm} }  
\newcommand{\rangind}[1]{\rangle\hspace{-0.2cm}{\phantom{\rangle}}_{#1}
\hspace{0.05cm}}
\begin{document}

\title{Exact response functions within the time-dependent Gutzwiller approach} 
\author{J.~B\"unemann$^{\rm a}$$^{\ast}$,  S.~Wasner$^{\rm b}$, E.~v.~Oelsen$^{\rm b}$, and G.~Seibold$^{\rm b}$, 
\thanks{$^\ast$Corresponding author. Email: buenemann@gmail.com\vspace{6pt}}\\
$^{\rm a}$Institut f\"ur Theoretische Physik, Leibniz Universit\"at, 30167 Hannover, Germany \\
$^{\rm b}$ Institut f\"ur Physik, BTU Cottbus, PO Box 101344, 03013 Cottbus,  
Germany}
\maketitle
 
\begin{abstract} 
 We  investigate the applicability of the two existing versions of a 
  time-dependent Gutzwiller approach (TDGA) 
 beyond the frequently used limit of infinite spatial dimensions. 
To this end, we study the two-particle response functions of a two-site
Hubbard model where we can compare the exact results and those derived
 from the TDGA. It turns out that only the more recently introduced version 
 of the  TDGA can be combined with a diagrammatic approach which allows for the
 evaluation of Gutzwiller wave functions in finite dimensions.
  For this TDGA method we derive the time-dependent Lagrangian for general 
 single-band Hubbard models.       
\end{abstract} 
 
\section{Introduction}\label{5tuu}
The discovery of high-temperature superconductivity in LaBaCuO by Bednorz 
and M\"uller \cite{bednorz1986} has led to 
 an enormous amount of theoretical and experimental work on unconventional 
 superconductivity (SC) in the past 25 years. There is now a wide agreement that 
 the Coulomb interaction among the conduction electrons of 
 such systems plays an important, or even the main role, in determining 
the properties of the SC order. 
 Unlike the conventional, i.e., phonon-mediated, pairing which can be understood 
 already on the level of a mean-field theory, a proper treatment of the Coulomb interaction
 requires genuine many-particle methods. Therefore, our theoretical understanding 
 of correlation-induced superconductivity is still far from satisfactory. 

For the theoretical investigation of cuprates, such as LaBaCuO, 
one  often considers a two-dimensional (single-band) Hubbard model
\begin{equation}\label{h2}
\hat{H}=\sum_{i\neq j} \sum_{\sigma=\uparrow,\downarrow}
t^{}_{i,j} \hcd_{i,\sigma}\hc_{j,\sigma}+
U\sum_i \hat{d}_{i}\;\;\;{\rm with} \;\;\;\hat{d}_{i}
=\hat{n}_{i,\uparrow}\hat{n}_{i,\downarrow}
\end{equation}
which describes the hopping of electrons with spin $\sigma$ between lattice 
 sites $i$ and $j$, and $U \hat{d}_{i}$ is the Coulomb interaction of electrons 
 on the same site.  Describing the superconducting ground states of this 
(still relatively simple) model, however, is already a  challenging task.
 We have recently introduced a variational approach, based on Gutzwiller 
 wave functions, which allows us to investigate the stability of 
 such states~\cite{buenemann2012a,buenemann2012b}. 
Unlike alternative methods which are based on the investigation 
 of finite (usually small) clusters, our method addresses the infinite 
 size system. This is important because, for an accurate description of 
superconductivity, one needs a sufficient resolution around the Fermi surface
 in momentum space. 

 The Gutzwiller wave function provides an approximation 
of the many-particle ground state and its properties. In order to go beyond the ground-state 
description there are two different formalisms which have been proposed in the literature.
 In Ref.~\cite{seibold2001}, Seibold and Lorenzana introduced a 
time-dependent Gutzwiller approach (TDGA)  which 
 was based on the assumption that the considered frequencies are small compared to typical atomic energies 
 (`antiadiabaticity assumption'). In a number of subsequent works on single-band 
models \cite{seibold2001,seibold2003,lorenzana2003,seibold2004,seibold2004b,lorenzana2005,seibold2005,seibold2006,seibold2007,seibold2008,seibold2008b}
 and (more recently) multi-band models \cite{buenemann2011b,buenemann2011f} it has
 been demonstrated that this 
 method provides a much more accurate description of low-energy excitations than, e.g., 
 a mean-field RPA calculation. Based on ideas by Schiro and Fabrizio 
\cite{schiro2010,schiro2011}, we have recently 
 derived an improved formulation of the time-dependent Gutzwiller approach  
which avoids the  antiadiabaticity assumption. This method
is therefore expected to extend the range of validity of the TDGA
to higher frequencies. \cite{buenemann2013a,buenemann2013b}.

  In all applications of the TDGA so far, expectation values
  have been evaluated  by means of the  `Gutzwiller approximation' (GA), see below.
  Certain phenomena, however,  cannot be understood within the GA, e.g., 
the superconductivity in a two-dimensional 
 Hubbard model. 
Therefore, the main question, which we will address in this work is 
how the TDGA can be used beyond this approximation, especially by means of the 
 diagrammatic approach which was introduced in Ref.~\cite{buenemann2012a}. 
 
Our work is organized as follows. In Sec.~\ref{5t} we introduce the 
 Gutzwiller wave function for the investigation of 
  Hubbard models. The main ideas of the two TDGA formulations 
 are summarised in Sec.~\ref{8t}. To assess the applicability 
 of these TDGA methods beyond the GA, it is  instructive to 
 consider a two-site Hubbard model where all expectation values
 for Gutzwiller wave functions are known exactly. We introduce 
 this model and useful notations in  Sec.~\ref{9t} and 
  investigate its two-particle properties by means 
 of the TDGA in  Sec.~\ref{10t}. Finally, in  Sec.~\ref{11t}
 we formulate the TDGA for lattice systems based on the 
 diagrammatic approach introduced in Ref.~\cite{buenemann2012a}.  
 A summary and an outlook close our presentation in Sec.~\ref{rf12}.

\section{Gutzwiller variational wave functions}\label{5t}
Gutzwiller wave functions (GWF) allow us to investigate the properties of 
 multi-band Hubbard models. They are defined as \cite{gutzwiller1963,buenemann1998}
\begin{equation}\label{1.3}
|\Psi_{\rm G}\rangle=\hat{P}_{\rm G}|\Psi_0\rangle=\prod_{i}\hat{P}_{i}|\Psi_0\rangle\;,
\end{equation}
where $|\Psi_0\rangle$ is an arbitrary normalised single-particle product state of the
system.  The form of the 
local Gutzwiller correlator depends on the  model one aims to investigate. 
 In this work, we will only consider single-band Hubbard models of the form~(\ref{h2}).
To study such models, Gutzwiller worked with the local correlation operator
\begin{equation}
\label{56}
 \hat{P}_{i}=1-(1-g_i)\hat{d}_{i}
\end{equation}
with variational parameters $g_i$ which connect the non-interacting ($g_i=1$) and  the 
`atomic' limit ($g_i\to 0$) \cite{gutzwiller1963}. An alternative definition of the local
 correlation operator is given by \cite{buenemann1998}
\begin{equation}
\label{57}
 \hat{P}_{i}=\sum_{\Gamma}\lambda_{i,\Gamma}|\Gamma \rangind{i}
\langind{i}  \Gamma |\;,
\end{equation}
where we introduced the four local basis states $\ket{\emptyset}_i$, $\ket{\sigma}_i$, 
$\ket{d}_i$ and corresponding variational parameters $\lambda_{i,\Gamma}$.
As we have shown in Ref.~\cite{buenemann2012a}, a diagrammatic evaluation of expectation values for
 Gutzwiller wave functions is significantly simplified if we write the four
 parameters $\lambda_{i,\Gamma}$ as
 \begin{equation}\label{89}
\lambda_{i,\emptyset}=\sqrt{1+n^0_{i,\uparrow}n^0_{i,\downarrow}x_i}\;\;,\;\;
\lambda_{i,\sigma}=\sqrt{1-\bar{n}^0_{i,\sigma}n^0_{i,\bar{\sigma}}x_i}\;\;,\;\;
\lambda_{i,d}=\sqrt{1+\bar{n}^0_{i,\uparrow}\bar{n}^0_{i,\downarrow}x_i}\;,
 \end{equation}
where the notations $\bar{\uparrow}=\downarrow$, $\bar{\downarrow}=\uparrow$, 
$n^0_{i,\sigma}=\langle \hat{n}_{i,\sigma} \rangle_{\Psi_0}$, $\bar{n}^0_{i,\sigma}\equiv 1-n^0_{i,\sigma}$  
and the variational 
 parameters $x_i$ have been introduced.
 
 Despite the different numbers of variational parameters, the
 correlation operators (\ref{56}) and (\ref{57}) (or  (\ref{57}) together with 
the transformation (\ref{89})) lead to the same  
 variational ground state as long as we consider 
 the single-particle state $|\Psi_0\rangle$ as a variational object. 
 One has to be more cautious, however, about the 
 proper choice of the correlation operator within the time-dependent
 Gutzwiller theory, as we will discuss in Sec.~\ref{10t}.
 A generalisation of the Gutzwiller Ansatz for  multi-band models is 
 straightforward~\cite{buenemann1998,buenemann2005,buenemann2012c} but needs 
not to be discussed for our considerations in this work.

 The evaluation of expectation values for Gutzwiller wave functions is a difficult many-particle 
 problem which can be solved exactly only in a few cases. For a one-dimensional (single-band) model, 
expectation values have been calculated for homogeneous para- and ferromagnetic 
states~\cite{metzner1987,kollar2002}. 
In the opposite 
 limit of infinite spatial dimensions it is possible to evaluate expectation values for general 
 multi-band models. The energy functional which arises in this limit is usually denoted as the 
 GA when it is applied to finite-dimensional systems. The diagrammatic
 approach introduced in Ref.~\cite{buenemann2012a} provides a systematic way to improve the GA. 

   \section{The time-dependent Gutzwiller approach (TDGA)}\label{8t}
In this section, we briefly summarise the two ways of formulating a TDGA.
The first, which in the following we term the `low frequency 
approximation'
(LFA), has been succesfully used previously within 
the GA.
The second is the fully time-dependent approach (FTDA) and will
turn out to be the appropriate method for generalizing
the calculation beyond the limit of infinite spatial dimensions.    
 
\subsection{LFA: the low-frequency approximation} \label{8tt}
 The expectation value of the Hamiltonian with respect to our Gutzwiller wave functions, 
\begin{equation}
\langle \hat{H} \rangle_{\Psi_{\rm G}}=E_{\rm G}(\{z_i \},\tilde{\rho})\,,
\end{equation}
 is a function of the (non-interacting) density matrix $\tilde{\rho}$ with the elements
\begin{equation}\label{r4}
\rho_{(j\sigma'),(i\sigma)}=\langle \hat{c}^{\dagger}_{i\sigma} \hat{c}^{}_{j\sigma'} \rangle_{\Psi_{0}}
\end{equation}
 and of the local variational parameters $z_i$ (i.e., $g_i$,  $\lambda_{i,\Gamma}$, or  $x_i$ in 
the single-band case).  We define the effective energy function
\begin{equation}\label{10.8}    
E(\tilde{\rho})\equiv \min_{\{z_i \}} E_{\rm G}(\{z_i \},\tilde{\rho})
\end{equation}
which, like the corresponding Hartree--Fock energy, only depends on the  
(non-interacting) density matrix $\tilde{\rho}$. 

For the study of response functions we add a time-dependent field
\begin{equation}\label{iae}
\hat{V}(t)=\sum_{i,j,\sigma}f_{(i\sigma),(j\sigma')}(t) \hat{c}^{\dagger}_{i\sigma} \hat{c}^{}_{j\sigma'} 
\end{equation}
 to 
 the Hamiltonian $\hat{H}$ of our system.  
In the time-dependent  Hartree--Fock approximation (also denoted as the 
`random-phase approximation') the time dependence of  $\tilde{\rho}$ is given by 
 the equation of motion
\begin{equation}\label{10.810}    
 {\rm i}  \dot{\tilde{\rho}}(t)=
[\tilde{h}(\tilde{\rho}(t))+\tilde{f}(t),\tilde{\rho}(t)]
\end{equation} 
where the elements of the Hamilton matrix  $\tilde{h}$ are given by the 
derivative 
 of the Hartree--Fock energy  with respect to the matrix elements of 
$\tilde{\rho}$. It is the main idea of the time-dependent Gutzwiller 
approach, as
 introduced in Ref.~\cite{seibold2001}, to evaluate Eq.~(\ref{10.810})
 with a  Hamilton matrix  $\tilde{h}$ which is derived from the effective 
 energy function~(\ref{10.8}),
\begin{equation}\label{10.9}   
h_{(i\sigma), (j\sigma') }=\frac{\partial}{\partial  \rho_{(j\sigma'),(i\sigma)}}E(\tilde{\rho})\;.
\end{equation} 
%\pacs{} to gallery 
The use of the effective energy function in Eqs.~(\ref{10.8}), (\ref{10.810}) has been denoted as the 
 `antiadiabaticity assumption' in previous works. Physically, it is based on the idea 
that the time scale
 of `atomic' fluctuations, described by the parameters $z_i$,  will be short compared to 
 the externally induced fluctuations of $\tilde{\rho}(t)$. Obviously, this approximation 
is justifiable only for the study of low-frequency excitations.     

Within the GA, the further evaluation of Eqs.~(\ref{10.8}), (\ref{10.810}),  (\ref{10.9}) 
for the calculation of two-particle 
response functions has been discussed in great detail in previous 
work \cite{seibold2001,seibold2003,lorenzana2003,seibold2004,seibold2004b,lorenzana2005,seibold2005,seibold2006,seibold2007,seibold2008,seibold2008b} and shall not be repeated  here. We will analyse these equations for the two-site Hubbard
 model in Sec.~\ref{10tt}. 

  \subsection{FTDA: The fully time dependent approach} \label{8ttt}
 The Schr\"odinger equation for a general time-dependent
Hamiltonian $\hat{H}(t)$
 ($\hbar=1$),
%\begin{equation}\label{tds}
%\rmi \frac{\partial}{\partial t}\ket{\Phi(t)}
%=\hat{H(t)}_t\ket{\Phi(t)}
%\end{equation}
can be obtained by requesting that the action 
\begin{equation}
  \label{eq:action}
  S=\int {\rm d}t L(t)
\end{equation}
is stationary with respect to variations of the wave function. 
It is usually convenient to perform this variation based
on a real Lagrangian~\cite{kramers1981} 
\begin{equation}
  \label{eq:lagran}
  L(t)=\frac{i}{2}\frac{\langle \Psi|\dot{\Psi}\rangle
-\langle \dot{\Psi}|\Psi\rangle} 
{ \langle \Psi | \Psi \rangle}
- \frac{\langle \Psi|\hat{H}|\Psi\rangle} 
{ \langle \Psi | \Psi \rangle}\equiv  L^{(1)}+L^{(2)}\;.
%\bra{\Phi(t)}i\partial_t-H(t)\ket{\Phi(t)}.
\end{equation}
If one restricts the wave-function $\ket{\Psi(\lbrace z_i \rbrace,t)}$ to a 
certain trial form, depending on (in general complex) functions 
$ z_i$, the differential equations
\begin{equation}\label{76}
\frac{\rm d}{{\rm d}t} \frac{\partial L}{\partial \dot{z}^{(*)}_i}
-\frac{\partial L}{\partial z^{(*)}_i}=0
\end{equation}
provide an approximation for
the exact time evolution. Note that the exact solution is reproduced by solving 
 Eqs.~(\ref{76}) if the former can be written in the form of the Ansatz   
wave-function $\ket{\Psi(\lbrace z \rbrace,t)}$. Moreover, the Ritz' variational principle is
 recovered if the Hamiltonian and, consequently, the parameters $z_i$ are time independent. 

Within the time-dependent Gutzwiller theory our obvious choice for an Ansatz wave function 
 is of the form~(\ref{1.3}) where both, $\ket{\Psi_0}$ and   $\hat{P}_{\rm G}$, may be time dependent.
 The evaluation of the differential equations~(\ref{76}) then requires the calculation of 
 \begin{equation}\label{79}
 \frac{\langle \Psi_{\rm G}|\hat{H}|\Psi_{\rm G}\rangle} 
{ \langle \Psi_{\rm G} | \Psi_{\rm G} \rangle}\;\;\;{\rm and} \;\;\;
\frac{\langle \Psi_{\rm G}|\dot{\Psi}_{\rm G}\rangle}
{ \langle \Psi_{\rm G} | \Psi_{\rm G} \rangle}
  \end{equation}
which, again, constitutes a difficult many-particle problem that cannot be solved in general. 
 In Ref.~\cite{buenemann2013a}, we have derived the differential 
equations~(\ref{76}) for general multi-band models 
by evaluating~(\ref{79}) in the limit of 
 infinite spatial dimensions. The exact evaluation of Eqs.~(\ref{76}) for the
 case of a two-site Hubbard model will be discussed in Sec.~\ref{10ttt}.
  Finally, in  Sec.~\ref{11t} we derive the differential equations~(\ref{76}) for
 general Hubbard models based on the diagrammatic approach introduced in Ref.~\cite{buenemann2012a}.

\section{The two-site Hubbard model}\label{9t}
A general two-electron state in the ($S_z=0$) Hilbert space of the two-site Hubbard model ($t\ge 0$)
\begin{equation}\label{2.150}
\hat{H}_{2{\rm s}}=-t\sum_{\sigma}\left(\hcd_{1,\sigma}\hc_{2,\sigma}
+\hcd_{2,\sigma}\hc_{1,\sigma}\right)+
U\sum^2_{i=1}\hat{n}_{i,\uparrow}\hat{n}_{i,\downarrow}
\end{equation}
 has the form
 \begin{equation}\label{2.15}
\ket{\Phi}=\alpha_1\ket{d,\emptyset}+\alpha_2\ket{\emptyset,d}+
\alpha_3\ket{\uparrow,\downarrow}+\alpha_4\ket{\downarrow,\uparrow}
 \end{equation}
where we introduced the four basis states 
 \begin{equation}
\ket{d,\emptyset}\equiv \hcd_{1,\uparrow}\hcd_{1,\downarrow}\ket{0}\;,\;
\ket{\emptyset,d}\equiv \hcd_{2,\uparrow}\hcd_{2,\downarrow}\ket{0}\;,\;
\ket{\uparrow,\downarrow}=\hcd_{1,\uparrow}\hcd_{2,\downarrow}\ket{0}\;,
\ket{\downarrow,\uparrow}=\hcd_{2,\uparrow}\hcd_{1,\downarrow}\ket{0}\;,
 \end{equation}
and (complex) coefficients $\alpha_i$. For the ground state, one finds $\alpha_1=\alpha_2\equiv \alpha_d$ and 
  $\alpha_4=\alpha_3\equiv -\alpha_{\rm s}$ with a ground state energy
  $E_0=(U-\sqrt{U^2+16t^2})/2$. In the non-interacting limit, these results
 reduce to 
 $\alpha_d=\alpha_{\rm s}=1/2$ and  $E_0=-2t$.

Any state of the form~(\ref{2.15}) can be written as a Gutzwiller 
 wave function, no matter which form  of the local correlation operator we 
 choose. We demonstrate this for the correlation 
operator~(\ref{56}). In a general state $\ket{\Psi_0}$, we create
 two particles described by the operators
\begin{equation}\label{9o}
\hhd_{\sigma} \equiv e^{{\rm i}\phi_{\sigma}}  u_{\sigma}\hcd_{1,\sigma} + v_{\sigma} \hcd_{2,\sigma} \;  \; \;
(\sigma=\uparrow,\downarrow)\;,
\end{equation}
where $v_{\sigma}\equiv \sqrt{1- u_{\sigma}^2}$.  
 The variational parameters in~(\ref{56}) are also 
 complex numbers in the following considerations and will be written 
as $g_i=\bar{g}_ie^{{\rm i}\kappa_i}$ with $ \bar{g}_i, \kappa_i\in 
\mathcal{R}$. With this expression for $g_i$ and Eq.~(\ref{9o}), a general Gutzwiller wave function for the 
 two-site Hubbard model has the form 
  \begin{equation}\label{2.15g}
\ket{\Psi_{\rm G}}=\bar{g}_1e^{{\rm i}(\kappa_1+\phi_{\uparrow}+\phi_{\downarrow})}u_{\uparrow} u_{\downarrow} \ket{d,\emptyset}
+ \bar{g}_2e^{{\rm i}\kappa_2} v_{\uparrow} v_{\downarrow}\ket{\emptyset,d}+
e^{{\rm i}\phi_{\uparrow}} u_{\uparrow} v_{\downarrow} \ket{\uparrow,\downarrow}
+ e^{{\rm i}\phi_{\downarrow}} v_{\uparrow} u_{\downarrow} \ket{\downarrow,\uparrow}.
 \end{equation} 
We write the coefficients in~(\ref{2.15}) as $\alpha_i=\bar{\alpha}_ie^{{\rm i}\beta_i}$ and chose 
all phases such that $u_{\sigma},\bar{g}_i,\bar{\alpha}_i>0$. Then the comparison of the phases 
 in ~(\ref{2.15})
 and~(\ref{2.15g}) yields
  \begin{equation}
\phi_{\downarrow}=\beta_4\;, \;\;\phi_{\uparrow}=\beta_3\;, \;\;\kappa_2=\beta_2\;, \;\;
\kappa_1=\beta_1-\beta_3-\beta_4\;.
  \end{equation} 
Instead of normalising both states~(\ref{2.15})
 and~(\ref{2.15g}), it is easier (and equivalent)  to set $\bar{\alpha}_4=1$ and 
 ensure the same for the corresponding coefficient in $\ket{\Psi_{\rm G}}$ 
by dividing~(\ref{2.15g}) by $v_{\uparrow} u_{\downarrow} $.
 A comparison of the three remaining coefficients then  leads to the equations
\begin{equation}
\bar{g}_1u_{\uparrow} =\bar{\alpha}_1 v_{\uparrow}\;, \;\;
\bar{g}_2 v_{\downarrow}=\bar{\alpha}_2u_{\downarrow}\;, \;\;
u_{\uparrow} v_{\downarrow}=\bar{\alpha}_3   u_{\downarrow} v_{\uparrow}          \;, 
  \end{equation} 
 which have the solution 
 \begin{equation}\label{hd}
\bar{g}_1=\frac{\bar{\alpha}_1\sqrt{1-u^2_{\downarrow}}}{\bar{\alpha}_3u_{\downarrow}}\;, \;\;
\bar{g}_2=\frac{\bar{\alpha}_2u_{\downarrow}}{\sqrt{1-u^2_{\downarrow}}}\;, \;\;
u_{\uparrow}=\frac{\bar{\alpha}_3u_{\downarrow}}{\sqrt{1-u^2_{\downarrow}(1-\bar{\alpha}^2_3)}}\;.
 \end{equation} 
With this result, we have demonstrated that each state of the form~(\ref{2.15}) can be written  
 as a Gutzwiller wave function with the correlation operator~(\ref{56}). The same can be shown 
 for the other representations of the correlation operators in Sec.~\ref{5t}. Note that 
 the parameters in the correlation operators are not uniquely
 defined by the correlated state. For example, any choice of $u_{\downarrow}$ in~(\ref{hd}) 
  leads to the same Gutzwiller wave function after normalisation. This ambiguity
 will turn out to be the main obstacle in our application of the LFA, see below.  

 \section{The TDGA for the two-site Hubbard model}\label{10t}
As we will show in this section,  one can gain valuable insight into the 
applicability of the time-dependent approach beyond the Gutzwiller 
approximation  by a comparison 
 with the exact results for a two-site Hubbard model. We shall discuss the 
low frequency (LFA) and fully time-dependent approach (FTDA)  
 separately in the following two sections. 
 \subsection{LFA}\label{10tt}
The LFA requires the calculation of the effective energy function~(\ref{10.8}). 
 If we work, e.g., with the correlation operator~(\ref{56}) we need to minimise the 
 energy with respect to $g_1$,  $g_2$ which gives us an effective energy 
$E(u_{\uparrow},u_{\downarrow})$ as a function of the two parameters $u_{\sigma}$  
 that  determine $\ket{\Psi_0}$. This effective energy  has then 
 to be used in the equations of motion~(\ref{10.810}), (\ref{10.9}).
To demonstrate the substantial difficulties of this approach
 we consider the case of pure charge fluctuations where 
\begin{eqnarray}
&&\bar{\alpha}_3=\bar{\alpha}_4,\;\; \beta_3=\beta_4=-\pi,\;\;\beta_1=\beta_2=0\;\;\;\;{\rm in\; (\ref{2.15} )}\;,\\
&&\phi_{\uparrow}=\phi_{\downarrow}=-\pi,\;\;  u_{\uparrow}=u_{\downarrow}\equiv u, \;\; \kappa_i=0\;\;\;\;{\rm in\; (\ref{2.15g})}\;.
\end{eqnarray}
 In this case, the ground state (with  $\bar{\alpha}_3=\bar{\alpha}_4=1, 
\bar{\alpha}_1\equiv \bar{\alpha}^{\rm gs}_1, \bar{\alpha}_2\equiv \bar{\alpha}^{\rm gs}_2 $) is recovered for 
 each value of $u$ by setting $\bar{g}^2_{1}=(1-u^2)/u^2\bar{\alpha}^{\rm gs}_1 $ and $\bar{g}^2_{2}=u^2/(1-u^2)
\bar{\alpha}^{\rm gs}_2$, see 
Eqs.~(\ref{hd}). 
 As a consequence, the effective energy function (equivalent to the exact ground state energy), 
\begin{equation}
E(\tilde{\rho})=E(u_{\uparrow},u_{\downarrow})=E(u)=E_0=(U-\sqrt{U^2+16t^2})/2\,,
\end{equation}
is a  constant,  i.e.,  independent of the density matrix.
This energy expression obviously leads to unphysical results 
 if we evaluate 
the equations of motion~(\ref{10.810}), (\ref{10.9}) and calculate, e.g., the 
 charge susceptibility $\chi_{\rm c}(\omega)\equiv
 \langle\! \langle \hat{n}_1;  \hat{n}_2 \rangle\!\rangle_{\omega}$ (with
 $\hat{n}_i\equiv \hat{n}_{i,\uparrow}+\hat{n}_{i,\downarrow}$).  

The problem with the correlation operator~(\ref{56}) stems from the fact that a fluctuation
 imposed on the state $\ket{\Psi_0}$ can be fully reversed by a proper choice of the parameters 
$g_1$,  $g_2$. Hence, the LFA does not even lead to the correct result in the non-interacting 
limit ($U=0$). Since~(\ref{56}) is a special version of the correlation operator~(\ref{57}), 
the latter runs into the same problem. Only if we add the transformation~(\ref{89}), 
the  correlation operator~(\ref{57}) leads to a  meaningful effective energy function
$E(\tilde{\rho})$ which, however, still does not give the exact response functions, see below.   

The problems which we encounter with the correlation operators~(\ref{56}), (\ref{57})
 are, at first, surprising because the LFA has been 
 applied successfully in a number of studies which were based on the energy 
function of the GA.
  Even a simple RPA calculation reproduces the correct response functions in 
 the non-interacting limit. The reason why the GA leads to sensible results lies in a technical 
 aspect of that approach whose physical consequences within the LFA have so far been overlooked. 
In order to evaluate
 expectation values in the limit of infinite dimensions analytically it is most convenient to work
 with the more general correlation operator~(\ref{57}) and to impose constraints which, for our two-site 
 model, have  the form
 \begin{equation}\label{set}      
\langle \hat{n}_{i,\sigma}  \rangle_{\Psi_0}=  \langle \hat{n}_{i,\sigma}  \rangle_{\Psi_{\rm G}}\;.
 \end{equation}
 As long as we are only interested in the variational ground state, these constraints 
 do not change the physics, i.e., they just specify a particular representation of 
$\ket{\Psi_{\rm G}}$  within the variational freedom contained in the correlation 
operator~(\ref{56}). 
In the context of the LFA, however, the constraints have significant 
consequences  because they ensure that fluctuations imposed on $\ket{\Psi_0}$ are 
 automatically also imposed on  $\ket{\Psi_{\rm G}}$. In this way, one avoids the difficulties
 which we observed for the correlation operator~(\ref{56}). 
An obvious remedy in our attempt to formulate the  LFA for finite-dimensional systems
 (i.e., beyond the GA) is therefore to use these  constraints  here as well. Obeying 
 the constraints, however, requires the use of the  correlation operator~(\ref{57}) because
 with  (\ref{56}) (or~(\ref{57}) together with the transformation~(\ref{89})) there 
 are no variational parameters left in the minimisation~(\ref{10.8}). 

We have calculated the charge susceptibility $\chi_{\rm c}(\omega)$ and the spin susceptibility
 $\chi_{\rm s}(\omega)\equiv \langle\! \langle \hat{S}^{z}_1;\hat{S}^{z}_2  \rangle \!\rangle_{\omega}$
  (with $ \hat{S}^{z}_i=(\hat{n}_{i,\uparrow}-\hat{n}_{i,\downarrow})/2$) with the three available 
 LFA methods, i.e., we use the correlation operator~(\ref{57}) i) with the 
constraint~(\ref{set}), 
 ii) with the transformation~(\ref{89}) 
where both i) and
ii) are based on the exact evaluation of the energy functional. Finally, 
and iii), we use the correlation operator~(\ref{57}) and consider the 
 energy function from the GA.  

\begin{figure}
  \begin{center}
    \subfigure[]{
      % Ersetzungen fuer linke Grafik
      \psfrag{exakt, lambda}{\hspace*{1.1cm}exact, $\lambda$}
      \psfrag{x}{$x$}
      \psfrag{GA}{GA}
      \psfrag{RPA}{RPA}
      \psfrag{E/|t|}{$E_{\rm S}/|t|$}
      \psfrag{U/8|t|}{$U/8|t|$}
      \resizebox*{7cm}{!}{\includegraphics{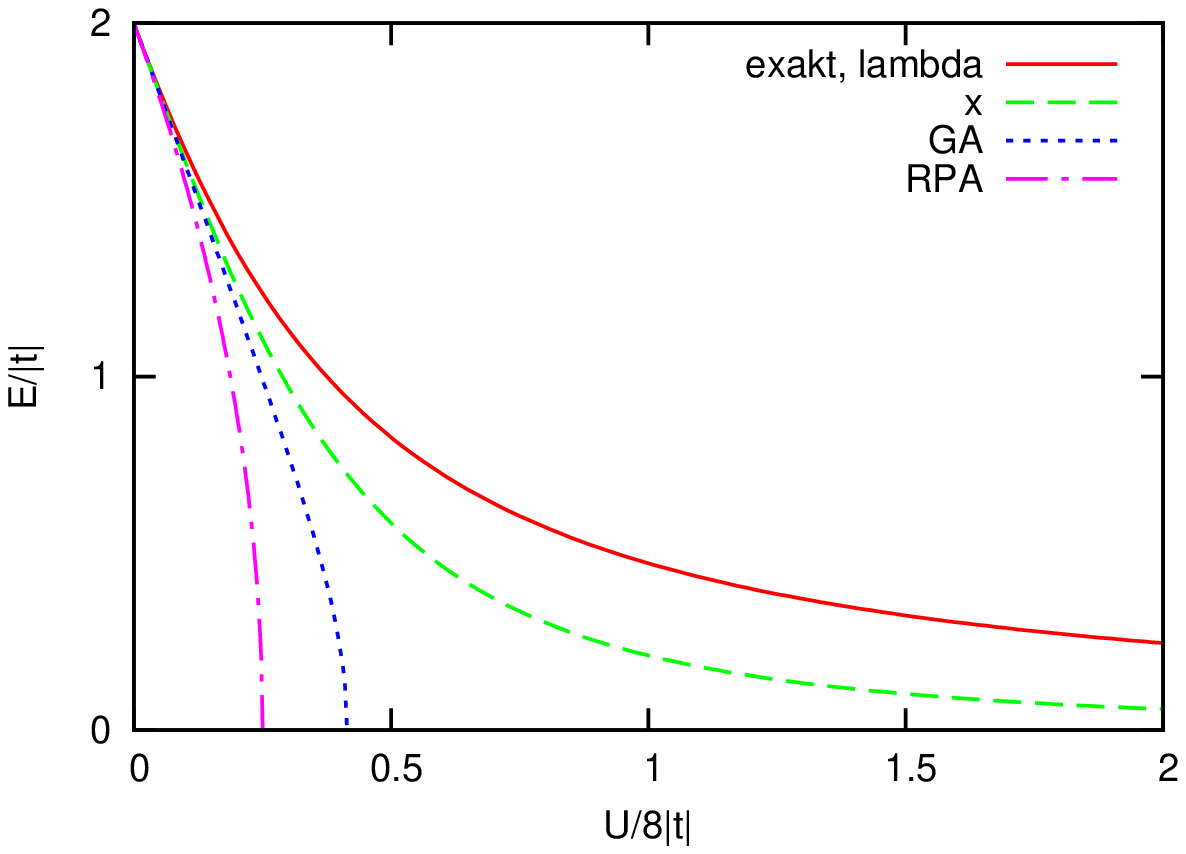}}}%
    \subfigure[]{
      % Ersetzungen fuer rechte Grafik
      \psfrag{exakt, lambda}{\hspace*{1.1cm}exact, $\lambda$}
      \psfrag{x}{$x$}
      \psfrag{GA}{GA}
      \psfrag{RPA}{RPA}
      \psfrag{E/|t|}{$E_{\rm c}/|t|$}
      \psfrag{U/8|t|}{$U/8|t|$}
      \resizebox*{7cm}{!}{\includegraphics{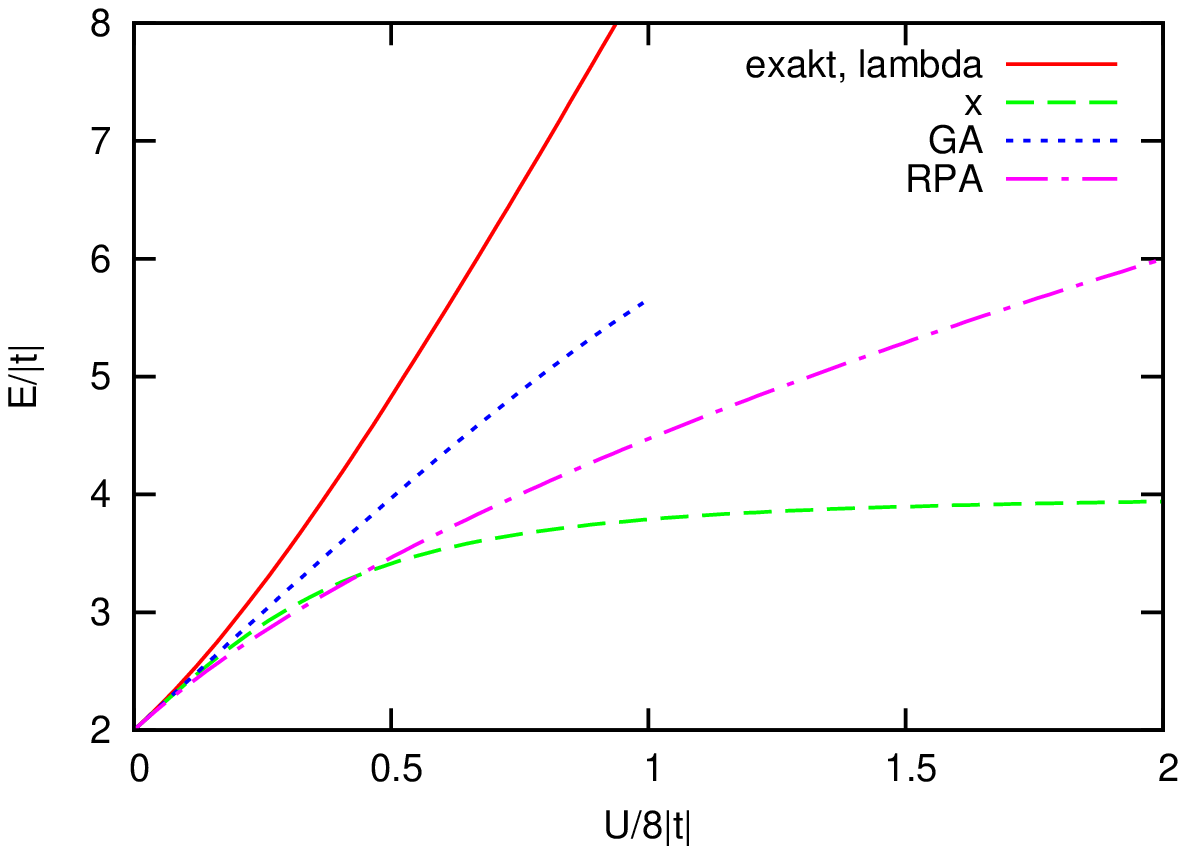}}}%
    \caption{The excitation energies of the spin (a) and the charge susceptibility (b)
      calculated with: the operator~(\ref{57}) and
      the constraint~(\ref{set}) ($\lambda$);
      the operator~(\ref{57}) and the transformation~(\ref{89}) ($x$); the
      Gutzwiller approximation (GA),  the random-phase approximation (RPA).}%
    \label{sample-figure}
  \end{center}
\end{figure}

As in any finite system, the imaginary parts of  $\chi_{\rm s/c}$ 
 have $\delta$-like peaks at certain excitation energies of the system. It turns out that 
the exact solution of our 
 two-site model has only one such energy for each of the two susceptibilities.  These are 
  \begin{equation}\label{set9}      
E^{\rm exact}_{\rm c}=(U+\sqrt{U^2+16t^2})\;\;\;{\rm and}\;\;\;E^{\rm exact}_{\rm s}=(-U+\sqrt{U^2+16t^2})\;.
 \end{equation} 
In Fig.~\ref{sample-figure}  we show these two excitation energies as well as 
their values 
in the Gutzwiller methods, mentioned
 above, and in the RPA (based on the Hartree--Fock approximation). 
The Gutzwiller method which employs the constraint~(\ref{set}) gives the 
exact results for the excitation energies. All other methods constitute approximations of 
 varying accuracy.

 In the GA and the RPA the excitation energies $E_{\rm s}$ go to zero at finite values
 of $U$ ($U^{\rm RPA}_{\rm crit}=2t$, $U^{\rm GA}_{\rm crit}=8(\sqrt{2}-1)t\approx 3.3t$).  At these points, the 
 GA and the Hartree--Fock 
 energy function predict a spurious transition to an antiferromagnetic N{\'e}el state. This 
 transition is absent if we work with the transformation~(\ref{89}) and the excitation energy
 is generally closer to the exact result for all $U$ than those calculated with the two other methods. 
  In contrast, the best approximation for the  excitation energy $E_{\rm c}$  is the  
  GA which, however, can only be applied up to the Brinkmann--Rice transition $U=U_{\rm BR}=8|t|$ where 
 the particles localise  in that approach ($\langle \hat{d}_i \rangle_{\Psi_{\rm G}}=0$). 

In summary, the LFA forces us to work with the constraint~(\ref{set}) if we want to 
 recover the exact results for the susceptibilities $\chi_{\rm c}(\omega)$ and 
 $\chi_{\rm s}(\omega)$ of the two-site Hubbard model based on the exact evaluation 
of $\ket{\Psi_{\rm G}}$. This prohibits us from using  the 
 transformation~(\ref{89}) which, however, is an indispensable element of our diagrammatic
 evaluation of expectation values beyond the GA. Therefore, we must conclude, that the 
 LFA is only of limited use if we aim to improve our calculation of susceptibilities 
  in finite dimensions.
\subsection{FTDA}\label{10ttt}
In Sec.~\ref{9t}, we have demonstrated that all states of the two-site Hubbard model
 can be written as a Gutzwiller wave function for each choice of the correlation 
 operator. Therefore, a time-dependent Gutzwiller Ansatz in the Langrangian~(\ref{eq:lagran}) 
  must reproduce the exact solution of the time-dependent Schr\"odinger equation for
 any time-dependent perturbation. This implies that for the susceptibilities which derive 
 from perturbations of the form~(\ref{iae}) the exact results will also be recovered.

Note that the differential equations for the parameters $z_i$ in the Gutzwiller wave
 function (e.g., $x_1, x_2$) are not necessarily linear and therefore 
  much more complicated than the (linear) Schr\"odinger equation (e.g., for 
the coefficients  $\alpha_i$ in (\ref{2.15})). 
  For the calculation of 
 response functions, however, we will always expand the differential 
 equations around the ground-state values of $z_i$  to linear order and,
 in this way, end up with linear equations.

From the study of the differential 
 equations for a two-site Hubbard model we do not learn much about the
 corresponding lattice problem, see below. Therefore, we conclude this section
 with the summary that all forms of correlation operators are, in principle,  suitable 
 for an improved calculation of two-particle excitation within the 
 FTDA method.  
 This holds, in particular, for the operator~(\ref{57}) with the  
 transformation~(\ref{89}) which allows us to formulate an efficient diagrammatic 
 evaluation of expectation values for finite-dimensional systems. In the following section, we will 
therefore derive  the general FTDA  equations for lattice systems.
 
\section{The time-dependent Gutzwiller theory for finite dimensional systems}\label{11t}
The transformation~(\ref{89}) ensures that  
\begin{equation}\label{pp}      
\hat{P}_i^{\dagger} \hat{P}_i^{} =1+x_i\hat{d}^{\rm HF}_i \;\;\;,\;\;\;
\hat{d}^{\rm HF}_i\equiv \hat{n}^{\rm HF}_{i,\uparrow}\hat{n}^{\rm HF}_{i,\downarrow}
\;\;\;,\;\;\;
\hat{n}_{i,\sigma}^{\rm HF}\equiv \hat{n}_{i,\sigma}-n^0_{i,\sigma}
 \end{equation}
 is obeyed by the correlation operator~(\ref{57}). As we have shown in detail in 
Ref~\cite{buenemann2012a}, this form of  
$\hat{P}_i^{\dagger} \hat{P}_i^{}$ allows for a very efficient diagrammatic expansion of 
 expectation values. Based on the same diagrammatic expansion 
 we will now derive the form of the Lagrangian~(\ref{eq:lagran}) for our Gutzwiller
 wave functions. For all details on the diagrammatic method we refer the reader to 
Ref.~\cite{buenemann2012a}.

A general time-dependent state $\ket{\Psi_0}$ has the form
\begin{equation}
   |\Psi_{0}(t)\rangle=\prod_{\gamma}[\hat{h}^{\dagger}_{\gamma}(t)]^{n_{\gamma}}\ket{\rm vac} 
\;\;\;{\rm with}\;\;\;
\hat{h}^{\dagger}_{\gamma}(t)=\sum_{i,\sigma}u_{(i\sigma),\gamma}(t)\hat{c}^{\dagger}_{i,\sigma}\;.
 \end{equation}
Here, $n_{\gamma}\in \{0,1 \}$ determines which of the single 
 particle states $\ket{\gamma(t)}$, described by the operators $\hat{h}^{\dagger}_{\gamma}$
are occupied and $u_{(i\sigma),\gamma}(t)$ is a (time-dependent) unitary transformation.
The time dependent parameters $\lambda_{i,\Gamma}(t)$ in~(\ref{57}) are written as
 \begin{equation}
\lambda_{i,\Gamma}(t)=e^{{\rm i}\varphi_{i,\Gamma}(t)}\sqrt{1+\Theta_{i,\Gamma}(t)x_i(t)}\;,
\end{equation}
where $\Theta_{i,\Gamma}(t)$ is given by the corresponding coefficients in~(\ref{89}), e.g., 
  $\Theta_{i,\emptyset}(t)=n^0_{i,\uparrow}(t)n^0_{i,\downarrow}(t)$. Note that the local expectation 
 values $n^0_{i,\sigma}$ are, like all elements of the non-interacting density matrix~(\ref{r4}),
 given as functions of $u_{(i\sigma),\gamma}(t)$, 
 \begin{equation}
n^0_{i,\sigma}(t)=\sum_{\gamma}n_{\gamma}|u_{i\sigma,\gamma}(t)|^2\;.
\end{equation}
 After having introduced all relevant time-dependent quantities, we will drop
 the explicit time dependence in the following considerations.

For the first term $L^{(1)}$ in~(\ref{eq:lagran}) we need to calculate 
 \begin{equation}\label{poa}
\langle \Psi_{\rm G} | \dot{\Psi}_{\rm G}  \rangle=\langle \Psi_{0} | 
\hat{P}^{\dagger}_{\rm G}  \hat{P}_{\rm G} ^{}|\dot{\Psi}_{0}  \rangle+
\langle \Psi_{0} | \hat{P}_{\rm G} ^{\dagger} \dot{\hat{P}}_{\rm G}| \Psi_{0}  \rangle\;.
\end{equation}
  The time derivative of $\ket{\Psi_{0}}$ leads to
 \begin{equation}\label{poa9}
\frac{\langle \Psi_{0} | \hat{P}_{\rm G}^{\dagger} \hat{P}_{\rm G}^{}|\dot{\Psi}_{0}  
\rangle}{  \langle \Psi_{\rm G}  | \Psi_{\rm G}  \rangle}=
\sum_{\gamma,\gamma',i,\sigma} n_{\gamma}\dot{u}_{i\sigma,\gamma} u^*_{i\sigma,\gamma'}
\frac{\langle \Psi_{0} |
\hat{P}_{\rm G}^{\dagger} \hat{P}_{\rm G}^{}\hhd_{\gamma'} \hh_{\gamma}|  \Psi_{0}\rangle }
{\langle \Psi_{\rm G}  | \Psi_{\rm G}  \rangle}
 \;.
\end{equation}
Note that we cannot conclude at this stage that $\gamma'$ has to be equal to $\gamma$
 in~(\ref{poa9}) as we could do in a time-dependent Hartree Fock calculation 
 (with $\hat{P}_{\rm G}=1$) or in the limit of infinite dimensions, see below. 
 To evaluate~(\ref{poa9}) further we write it as
 \begin{equation}\label{poa9b}
(\ref{poa9})=\sum_{\gamma,i,j,\sigma} n_{\gamma}\dot{u}_{i\sigma,\gamma} u^*_{j\sigma,\gamma}
\frac{\langle \Psi_{0} |
\hat{P}_{\rm G}^{\dagger} \hat{P}_{\rm G}^{}\hcd_{i,\sigma} \hc_{j,\sigma}|  \Psi_{0}\rangle }
{\langle \Psi_{\rm G}  | \Psi_{\rm G}  \rangle}\equiv 
\sum_{\gamma,i,j,\sigma} n_{\gamma}\dot{u}_{i\sigma,\gamma} u^*_{j\sigma,\gamma}
R_{(i\sigma),(j\sigma)}
\end{equation}
where the expectation value $R_{(i\sigma),(j\sigma)}$ can now be calculated by means of the 
 diagrammatic method introduced in Ref.~\cite{buenemann2012a}. This leads to
  \begin{eqnarray}\nonumber
R_{(i\sigma),(j\sigma)}&=&\delta_{i,j}n^{c}_{i,\sigma}+(1-\delta_{i,j})
\Big[T^{(1),(1)}_{i\sigma,j\sigma}+
(1-n^0_{i,\sigma})x_iT^{(1),(3)}_{i\sigma,j\sigma}\\\label{wse}
&&-n^0_{j,\bar{\sigma}}x_jT^{(3),(1)}_{i\sigma,j\sigma}
-(1-n^0_{i,\sigma})n^0_{j,\bar{\sigma}}x_ix_j
T^{(3),(3)}_{i\sigma,j\sigma}\Big]
\end{eqnarray}
where 
\begin{equation}\label{iopr}
n^{c}_{i,\sigma}=\langle \hat{n}_{i,\sigma}  \rangle_{\Psi_{\rm G}}=n^0_{i,\sigma}+I^{(2)}_{\sigma}
+x_i(1-n^0_{i,\sigma})I^{(2)}_{\bar{\sigma}}+x_i(1-2n^0_{i,\sigma})I^{(4)}
\end{equation}
is the (correlated) local particle number. Expressions for the diagrammatic sums in~(\ref{wse})
 and~(\ref{iopr})
   have been derived in   
Ref.~\cite{buenemann2012a} and are given as
\begin{eqnarray}\label{sd}
I^{(4)[(2)]}_{[\sigma]}&\equiv&\sum_k\frac{1}{k!}\sum_{l_1,\ldots, l_k}
\bigl\langle 
\hat{d}^{\rm HF}_{i}[\hat{n}^{\rm HF}_{i,\sigma}]\prod_{m=1}^kx_{l_m}\hat{d}^{\rm HF}_{l_m}
\bigr\rangle^{\rm con}_{\Psi_0}
\label{eq:urtz}\; ,\\\label{df}
T_{(i\sigma),(j\sigma)}^{(1)[(3)],(1)[(3)]}(k)&\equiv&\sum_k\frac{1}{k!}
\sum_{l_1,\ldots,l_k}
\bigl\langle
[\hat{n}^{\rm HF}_{i,\bar{\sigma}}]
\hat{c}^{\dagger}_{i,\sigma}
[\hat{n}^{\rm HF}_{j,\bar{\sigma}}]
\hat{c}_{j,\sigma}^{\phantom{\dagger}}\prod_{m=1}^kx_{l_m}\hat{d}^{\rm HF}_{l_m}
\rangle^{\rm con}_{\Psi_0}\;.
\end{eqnarray}%
Here, $\langle \dots \rangle_{\Psi_0}^{ \rm con}$ indicates that only connected diagrams 
are to be kept after the application of Wick's theorem. Note that, in the limit 
 of infinite spatial dimensions (i.e., within the GA), 
we find $n^{c}_{i,\sigma}=n^0_{i,\sigma}$ and the only 
  non-zero diagram~(\ref{df}) is 
\begin{equation}\label{670}
T_{(i\sigma),(j\sigma)}^{(1)(1)}=\langle \hcd_{i,\sigma}  \hc_{j,\sigma}  \rangle_{\Psi_0}\;.
\end{equation}
With 
\begin{equation}
\sum_ju^*_{j\sigma,\gamma}\langle \hcd_{i,\sigma}  \hc_{j,\sigma}  \rangle_{\Psi_0}=u^*_{i\sigma,\gamma}
\end{equation}
we then recover the result
\begin{equation}
\frac{\langle \Psi_{0} | \hat{P}_{\rm G}^{\dagger} \hat{P}_{\rm G}^{}|\dot{\Psi}_{0}  
\rangle}{  \langle \Psi_{\rm G}  | \Psi_{\rm G}  \rangle}=\sum_{\gamma,i,\sigma} n_{\gamma}\dot{u}_{i\sigma,\gamma} u^*_{i\sigma,\gamma}
\end{equation}
which has been used in our previous work, Ref.~\cite{buenemann2013a}.

 For the second term in (\ref{poa}), we need to evaluate
\begin{equation}
\hat{P}_{\rm G} ^{\dagger} \dot{\hat{P}}_{\rm G}=\sum_i\Big (\prod_{j(\ne i)} \hat{P}_j^{\dagger}
\hat{P}_j
\Big)
\hat{P}_i ^{\dagger} \dot{\hat{P}}_{i}
\end{equation}
where 
\begin{equation}\label{skj}
\hat{P}_i ^{\dagger} \dot{\hat{P}}_{i}=\sum_{\Gamma}
\big[
 {\rm i}\dot{\varphi}_{i,\Gamma}
|\lambda_{i,\gamma}|^2\hat{m}_{i,\Gamma}-\frac{1}{2}(\dot{\Theta}_{i,\Gamma}x_i+
\Theta_{i,\Gamma}\dot{x}_i)\hat{m}_{i,\Gamma}\big]\;.
\end{equation}\label{skj2}
Note that $\Theta_{i,\Gamma}$ and $x_i$ are real numbers and, therefore, the 
 corresponding terms in~(\ref{skj}) do not enter our real Lagrangian~(\ref{eq:lagran}).
 Hence, its first part  is given as 
 \begin{equation}\label{hh}
L^{(1)}=\sum_{i,j}\Big[\frac{\rm i}{2} \sum_{\gamma,\sigma}n_{\gamma}
(\dot{u}_{i\sigma,\gamma} u^*_{j\sigma,\gamma}R_{(i,\sigma),(j,\sigma)}-{\rm h.c.})
-\delta_{i,j}\sum_{\Gamma} \dot{\varphi}_{i,\Gamma}  m_{i,\Gamma}  \Big]
\;,
\end{equation}
where we introduced the expectation value $m_{i,\Gamma} \equiv \langle \hat{m}_{i,\Gamma} \rangle_{\Psi_{\rm G}}$
 which can  be calculated diagrammatically. In infinite dimensions, as used in 
Ref.~\cite{buenemann2013a}, 
it has the simple form 
$m_{i,\Gamma}=|\lambda_{i,\Gamma}|^2\langle \hat{m}_{i,\Gamma} \rangle_{\Psi_0}$.

The Lagrangian~(\ref{hh}) is complemented by the expectation value of the Hamiltonian 
 in~(\ref{eq:lagran}).  Our diagrammatic approach leads to 
 \begin{eqnarray}\nonumber
L^{(2)}&=&-\sum_{i,j,\sigma}t_{i,j}\Big[q_{i,\sigma}q^*_{j,\sigma}T^{(1),(1)}_{i\sigma,j\sigma}+
q_{i,\sigma}\alpha^*_{j,\sigma}T^{(1),(3)}_{i\sigma,j\sigma}+\alpha_{i,\sigma}q^*_{j,\sigma}T^{(3),(1)}_{i\sigma,j\sigma}
+\alpha_{i,\sigma}\alpha^*_{j,\sigma}T^{(3),(3)}_{i\sigma,j\sigma}\Big]\\\label{00}
&&-U\sum_{i}m_{i,d}
\end{eqnarray}
where
 \begin{eqnarray}
q_{i,\sigma}&=&
\lambda^*_{d}\lambda_{\bar{\sigma}}n^0_{\bar{\sigma}}+\lambda^*_{\sigma}\lambda_{\emptyset}(1-n^0_{\bar{\sigma}})
\Big|_{i}=e^{-{\rm i}\chi_{\sigma}}(q_{\emptyset,\sigma}+q_{d,\sigma} e^{-{\rm i}\eta})\Big|_{i}\;,\\
\alpha_{i,\sigma}&=&\lambda^*_{d}\lambda_{\bar{\sigma}}-\lambda^*_{\sigma}\lambda_{\emptyset}\Big|_{i}
=e^{-{\rm i}\chi_{\sigma}}(\alpha_{\emptyset,\sigma}-\alpha_{d,\sigma} e^{-{\rm i}\eta})\Big|_{i}\;,
\end{eqnarray}
and 
\begin{align}
q_{\emptyset,\sigma}&\equiv|\lambda_{d}||\lambda_{\bar{\sigma}}|n^0_{\bar{\sigma}}\Big|_{i}\;\;,&
q_{d,\sigma}&\equiv|\lambda_{\sigma}||\lambda_{\emptyset}|(1-n^0_{\bar{\sigma}})\Big|_{i}\;\;,
\\
\alpha_{\emptyset,\sigma}&\equiv|\lambda_{d}||\lambda_{\bar{\sigma}}|\Big|_{i}\;\;,&
\alpha_{d,\sigma}&\equiv|\lambda_{\sigma}||\lambda_{\emptyset}|\Big|_{i}\;\;,\\
\chi_{\sigma}&\equiv\varphi_{\sigma}-\varphi_{\emptyset}\Big|_{i}\;\;,&
\eta&\equiv\varphi_{\uparrow}+\varphi_{\downarrow}-\varphi_{\emptyset}-\varphi_{d}\Big|_{i}\;.
\end{align}
As mentioned before, in infinite dimensions only the the kinetic energy 
diagram~(\ref{670}) is non-zero. 
 With this expression 
we recover 
 the Lagrangian derived in \cite{buenemann2013a}. As in that work, we have eliminated the 
phases  $\varphi_{i,\sigma}$,
$\varphi_{d,i}$ in favor of $\chi_{i,\sigma}$ and
$\eta_{i,\sigma}$. Note that after this elimination, $\varphi_{\emptyset,i}$ does not appear
anywhere in $L^{(2)}$ and therefore can be disregarded as
a dynamical variable. The Lagrangian $L^{(1)}$ then has the form
\begin{eqnarray}
L^{(1)}&=&-\sum_i m_{i,d}\dot \eta_i -
 \sum_{i,\sigma} n^{\rm c}_{i,\sigma}\dot\chi_{i,\sigma}
+\frac{\rm i}{2}
 \sum_{i,j,\sigma,\gamma}n_{\gamma}(\dot{u}_{i\sigma,\gamma} u^*_{j\sigma,\gamma}R_{(i,\sigma),(j,\sigma)}
-{\rm h.c.})\\\nonumber
&&-
\sum_{i,j}\Omega_{i,j}(t)
\Big(\sum_{\gamma}u^*_{i,\gamma}u_{j,\gamma}-1\Big)
\end{eqnarray}
where we have added a Lagrange-parameter term to ensure that 
$u_{i,\gamma}$ is unitary.  

With the Lagrangian~(\ref{eq:lagran}) derived, it is now a straightforward task to set up
 the differential equations for our dynamical variables
  $x_i$, $u_{i\sigma,\gamma}$, $\chi_{i,\sigma}$ and
$\eta_{i,\sigma}$. If evaluated around the ground-state values of 
 these properties, we are able to  calculate two-particle
 response functions. Technically, the most challenging part is the calculation of 
 first and second derivatives of the diagrams~(\ref{sd}) and~(\ref{df})
 with respect to the dynamical variables. Work on this numerical problem 
 is in progress and will be published
 elsewhere.

\section{Summary and Outlook}\label{rf12}
We have outlined a formalism which allows for the computation of
collective excitations on top of the {\it exact} Gutzwiller ground state,
i.e., beyond the Gutzwiller approximation corresponding to the 
limit of infinite spatial dimensions. We have outlined the approach
by means of the two-site Hubbard model where it reproduces the
exact excitation spectrum and have compared it with approximations used
earlier in this context. In future work the method can be used
in order to systematically improve the calculation of dynamical
correlation functions based on the Gutzwiller wave-function. Especially
interesting in this context is an investigation of the `Thouless criterion'
signaling the instability towards superconductivity since it is known
 \cite{seibold2008b,buenemann2012b}
that one has to go beyond the standard Gutzwiller approximation in order 
to stabilize SC order. Another obvious application is the study of
excitations within the one-dimensional Hubbard model. 
It will then be interesting to see to which extend characteristic
features of 1D correlated systems (e.g., spin-charge separation) 
are captured within the exact Gutzwiller correlations.

\section*{Acknowledgements}
We thank J. Lorenzana for helpful criticism.

\bibliographystyle{unsrt}
\bibliography{bib4}

\begin{thebibliography}{10}
\newcommand{\noopsort}[1]{}
\newcommand{\printfirst}[2]{#1}
\newcommand{\singleletter}[1]{#1}
\newcommand{\switchargs}[2]{#2#1}
\providecommand{\url}[1]{\normalfont{#1}}
\providecommand{\urlprefix}{Available at }

\bibitem{bednorz1986}
J.G. Bednorz and K.A. M{\"u}ller, Z. Phys. B 64 (1986), p. 189.

\bibitem{buenemann2012a}
J. B{\"u}nemann, T. Schickling, and F. Gebhard, Europhys. Lett. 98 (2012), p.
  27006.

\bibitem{buenemann2012b}
J. Kaczmarczyk, J.Spa{\l}ek, T. Schickling, and J. B{\"u}nemann, Phys. Rev. B 88
  (2013), p. 115127.

\bibitem{seibold2001}
G. Seibold and J. Lorenzana, Phys.~Rev.~Lett. 86 (2001), p. 2605.

\bibitem{seibold2003}
G. Seibold, F. Becca, and J. Lorenzana, Phys.~Rev.~B 67 (2003), p. 085108.

\bibitem{lorenzana2003}
J. Lorenzana and G. Seibold, Phys.~Rev.~Lett. 90 (2003), p. 066404.

\bibitem{seibold2004}
G. Seibold and J. Lorenzana, Phys.~Rev.~B 69 (2004), p. 134513.

\bibitem{seibold2004b}
G. Seibold, F. Becca, P. Rubin, and J. Lorenzana, Phys.~Rev.~B 69 (2004), p.
  155113.

\bibitem{lorenzana2005}
J. Lorenzana, G. Seibold, and R. Coldea, Phys.~Rev.~B 72 (2005), p. 224511.

\bibitem{seibold2005}
G. Seibold and J. Lorenzana, Phys.~Rev.~Lett. 94 (2005), p. 107006.

\bibitem{seibold2006}
G. Seibold and J. Lorenzana, Phys.~Rev.~B 73 (2006), p. 144515.

\bibitem{seibold2007}
G. Seibold and J. Lorenzana, Journal of Superconductivity and Novel Magnetism
  20 (2007), p. 619.

\bibitem{seibold2008}
G. Seibold, F. Becca, and J. Lorenzana, Phys.~Rev.~Lett. 100 (2008), p. 016405.

\bibitem{seibold2008b}
G. Seibold, F. Becca, and J. Lorenzana, Phys.~Rev.~B 78 (2008), p. 045114.

\bibitem{buenemann2011b}
E. v.  Oelsen, G. Seibold, and J. B{\"u}nemann, Phys. Rev. Lett. 107 (2011), p.
  076402.

\bibitem{buenemann2011f}
E. v.  Oelsen, G. Seibold, and J. B{\"u}nemann, New J. Phys. 10 (2011), p.
  113031.

\bibitem{schiro2010}
M. Schiro and M. Fabrizio, Phys.~Rev.~Lett. 105 (2010), p. 076401.

\bibitem{schiro2011}
M. Schiro and M. Fabrizio, Phys.~Rev.~B. 83 (2011), p. 165105.

\bibitem{buenemann2013a}
J. B{\"u}nemann, M. Capone, J. Lorenzana, and G. Seibold, New J. Phys. 10
  (2013), p. 053050.

\bibitem{buenemann2013b}
G. Seibold, J. B{\"u}nemann, and J. Lorenzana, J.~Supercond.~Nov.~Magn.
  (2013); doi=10.1007/s10948-013-2412-0.

\bibitem{gutzwiller1963}
M. Gutzwiller, Phys.~Rev.~Lett 10 (1963), p. 159.

\bibitem{buenemann1998}
J. B{\"u}nemann, W. Weber, and F. Gebhard, Phys.~Rev.~B 57 (1998), p. 6896.

\bibitem{buenemann2005}
J. B{\"u}nemann, F. Gebhard, and W. Weber, in \emph{Frontiers in Magnetic
  Materials}, A. Narlikar, ed., Springer, Berlin,  2005.

\bibitem{buenemann2012c}
J. B{\"u}nemann, F. Gebhard, T. Schickling, and W. Weber, physica status solidi
  (b) 249 (2012), p. 1282.

\bibitem{metzner1987}
W. Metzner and D. Vollhardt, Phys.~Rev.~Lett. 59 (1987), p. 121.

\bibitem{kollar2002}
M. Kollar and D. Vollhardt, Phys.~Rev.~B 65 (2002), p. 155121.

\bibitem{kramers1981}
P. Kramers and M. Saraceno, \emph{{Geometry of the Time-Dependent Variational
  Principle in Quantum Mechanics}}, Springer, {Berlin}, 1981.

\end{thebibliography}
%\bibliography{tPHMguide}

\end{document}